# AN AUTONOMOUS LONG RANGE MONITORING SYSTEM FOR EMERGENCY OPERATORS

Matteo Lanati[1], Davide Curone[1], Emanuele Lindo Secco[1], Giovanni Magenes[1,2]
and Paolo Gamba[1,3]

[1]European Centre for Training and research in Earthquake Engineering, Pavia, Italy
matteo.lanati@eucentre.it, davide.curone@eucentre.it,
emanuele.secco@eucentre.it

[2]Department of Computer Engineering and Systems Science, University of Pavia, Italy
giovanni.magenes@unipv.it

[3]Department of Electronic Engineering, University of Pavia, Italy
paolo.gamba@unipv.it

## ABSTRACT

*Miniaturization and portability of new electronics lead up to wearable devices embedded within garments: a European program called ProeTEX developed multi-purpose sensors integrated within emergency operators' garments in order to monitor their health state and the surrounding environment. This work deals with the development of an autonomous Long Range communication System (LRS), suitable to transmit data between operators' equipment and the local command post, where remote monitoring software is set up. The LRS infrastructure is based on Wi-Fi protocol and modular architecture. Field tests carried out on the developed prototype showed a high reliability in terms of correctly exchanged data and recovering capabilities in case of temporary disconnection, due to the operator's movements.*

## 1. INTRODUCTION

Recent progress on technology leads up to the realization of small and light electronics to be easily embedded within portable devices and even garments [1]; this miniaturization has involved the sensors, processors and communication modules in such a way that novel applications based on portable and wearable electronics have been realized. Portability have also introduced new scenarios like monitoring people's activity while carrying or dressing these systems: literature reports projects on monitoring individuals under different scenarios, like daily life, working activity, sports, fitness or rehabilitation [2-5]. A huge set of applications on elderly surveillance, health supervision, activity and context recognition is underway [6-8]. Some of these applications have been merely realized by transforming common portable devices into smart instruments, thanks to insertion of transducers - like accelerometers, sensors or microphones embedded within cell phones, notebooks, PDAs [9-11]. Similarly, sensors have been placed within sport garments and facilities: it is the case of GPS modules in watches, accelerometers on shoes, chest bands (for heart rate monitoring) of runners [12-14]. Another application leaded to the real-time monitoring of workers in harsh and dangerous environments, like emergency operators and soldiers: scientific literature reports several examples of wearable systems acquiring physiological, activity and environmental related data, aimed at improving the efficiency, coordination and safety under these conditions [2, 14-17]. Moreover, many projects can be cited about these applications: the SCIER project (Sensor & Computing

Infrastructure for Environmental Risks) [18], for example, is based on wireless sensor networks and on techniques of environmental engineering and modelling aimed at detecting and predicting natural hazards; the ARTEMIS project (Advanced Research Testbed for Medical Informatics) [16], developed by the US military's 21st Century Land Warrior Program and Defense Advance Research Projects Agency, worked on communication, sensorial and software systems for remote triage and emergency management of soldiers during combat. Furthermore the European Commission paid a great attention to these applications [19], by financing projects like WearIt@Work [20] and ProeTEX [14].

This work relates to the aforementioned ProeTEX project, aimed at supporting emergency operators - namely Civil Protection rescuers and fire-fighters - with equipments made of wearable sensor networks and communication-decision architectures. In the first section of this paper we will summarize the main characteristics of this equipment, as previously detailed in other literature [3, 21], whereas in the second part we will focus on the architecture of the communication system and decision-making support software.

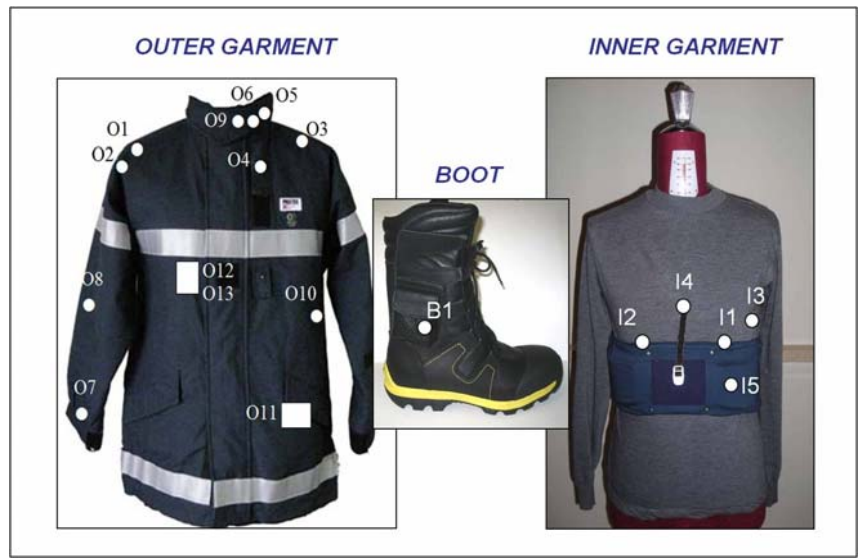

Figure 1. Outer Garment, Boot and Inner Garment of ProeTEX prototype. External temperature and heat flux sensors (O1 & O2), GPS module (O3), visual and acoustic alarm modules (O4 & O5), collar and wrist accelerometers (O6 & O7), textile motion sensor (O8), CO sensor (O9), Zig-Bee module, PEB, Wi-Fi modules and textile antennas (O10-O13). $CO_2$ sensor and Zig-Bee module housing (B1). Textile electrodes and piezo-electric sensor (for heart rate and breathing rate monitoring – I1 & I2), body temperature and $SpO_2$ sensors, vital signs board (I3-I5).

## 2. THE BODY SENSOR NETWORK

ProeTEX is a 4-year project begun in 2006: it foresaw the development of 3 prototypes of sensorized garments for emergency operators: each prototype is based on a T-shirt (called Inner Garment or IG), a jacket (Outer Garment or OG), plus a pair of boots; these units are endowed with sensors which monitor physiological, activity-related and environmental parameters; the IG integrates textile electrodes for heart rate monitoring, piezo-resistive and piezo-electric transducers for breathing rate monitoring, a body temperature sensor and a non invasive $SpO_2$ sensor to capture oxygen concentration within blood. The OG includes 2 tri-axial accelerometers, a GPS module, an external temperature sensor, acoustic and visual alarm modules (i.e. subsystems launching visual and acoustic warnings when one or more sensors detect possible dangerous conditions), a CO (carbon monoxide) sensor and a heat flux sensor. One pressure transducer plus a $CO_2$ (carbon dioxide) sensor are embedded in the two boots. An overview of the equipment is reported in the Figure 1. Each sensor is mounted with a low-power microcontroller, which samples the transducer signals and implements simple signal-processing routines to extract "high-level" features, updated at 1 Hz frequency. This distributed computing solution allows reducing the amount of data that each sensor's module has to output. For example, the tri-axial accelerometers data are locally sampled at 100 Hz, and then falls to the ground, body orientation [22] and movement's intensity [23] are online evaluated. Features of all sensors are real-time transmitted to a Professional Electronic Box (PEB) within the OG. Communications between OG and IG sensors are performed through RS485 bus (a part from the GPS module having a dedicated connection) and a Zig-Bee wireless link respectively. Data from boots sensors are interfaced with PEB through Zig-Bee wireless link too. A summary of these Body Area Network (BAN) sensors, connections and throughput is reported in Table I. More details on these sensors and garments and on the whole architecture can be found on [3].

PEB polls all the sensors with 1 Hz rate and aggregates all the outputs into a unique string. It also adds to the string some messages containing information related to the whole system status: device ID, internal batteries level of charge and timestamp of the data. At each second the ProeTEX equipment generates 390 bytes of data to be remotely transmitted. These data have to be wireless communicate to the intervention managers (i.e. officers overseeing the activity of the first line responders from a mobile command post in a safe place near the intervention area, according to Civil Protection agencies and fire-fighters current practice) by means of a customized developed Long Range communication System (LRS).

Table 1. Sensor module configuration – II prototype.

| sensor module | garment | function | PEB interface | data rate [bytes/sec] |
|---|---|---|---|---|
| piezo-electric sensor | IG | breathing rate | Zig-Bee | 24 |
| textile electrodes, piezo-resistive and body temp. sensors | IG | heart rate, breathing rate and body temperature | Zig-Bee | 68 |
| $SpO_2$ sensor | IG | oxygen saturation | Zig-Bee | 24 |
| 3D accelerometer n. 1 | OG | inactivity sensor | RS485 | 24 |
| 3D accelerometer n. 2 | OG | inactivity/fall sensor | RS485 | 24 |
| CO sensor | OG | carbon monoxide concentration | RS485 | 24 |
| external temp. sensor | OG | environmental temperature | RS485 | 16 |
| heat flux sensor | OG | heat flux across the jacket | RS485 | 16 |
| textile motion sensor | OG | inactivity sensor | RS485 | 24 |
| GPS module | OG | absolute position | Dedicated | 62 |
| $CO_2$ sensor | BOOT | carbon dioxide concentration | Zig-Bee | 20 |
| PEB related data | - | device ID, status, batteries, timestamp | - | 84 |

## 3. THE LONG RANGE COMMUNICATION SYSTEM

As soon as the data have been processed by sensor modules and collected by PEB, they are wireless transmitted to a remote PC of the intervention coordinator; this information exchange is based on a query-answer protocol, with queries driven by software on the PC and answers managed by the PEB. The system should guarantee robust and reliable communication link to helping the coordinator while managing the operators on the emergency field.

### 3.1. Requirements and specifications

According to ProeTEX aims, an ideal LRS should satisfy the following requirements and specifications:
- it should effectively cover the *distance* between the disaster area and the coordinator, normally 1-2 km far away from the team, with no possibility of visually inspect rescuers

- it must be *self-standing* in terms of communication infrastructure and power supply, since standard communication systems could have collapsed during the catastrophic events
- *modularity*: according to the extension of the disaster area and to the number of team and rescuers involved, the LRS network should be extendible and flexible
- *redundancy*: since the system should operate with sensitive data, robustness of the communication in response to possible losses of connection should be guaranteed too.

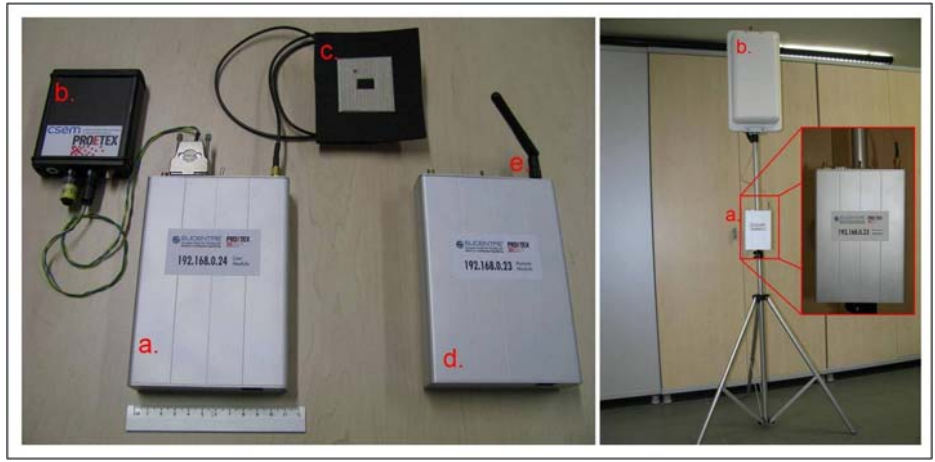

Figure 2. Left panel: the OP node (a), connected to the PEB (b) and textile antenna (c); the RT node (d) with the omni-directional antenna only (e); right panel: the tripod supporting the RT or RM node (a) with the directional antenna (b).

### 3.2. Choice of the communication protocol

Many protocols were considered to fit the requirements of this application: in particular, cellular systems such as GSM/GPRS or UMTS, TETRA, Wi-Fi and Wi-MAX are potential solutions for long range communication [24].

*Cellular systems* address mobility and power adaptation, provide data encryption, flexibility, large bandwidth and user authentication. Nevertheless, they need networks of base stations to correctly operate; unfortunately nets may collapse during interventions following a disaster or they could be unavailable due to high voice traffic. *TETRA* (Terrestrial Trunked Radio) is a standard of ETSI (European Telecommunications Standards Institute) for voice and data communication, allowing a large variety of transmission configurations: point-to-point, dispatch operation, talk-through, vehicle-mounted repeater are possible configurations. It operates in circuit mode (simultaneous transmission of voice and data within a channel allocated for the whole call duration) or direct mode (direct mobile-to-mobile communications without any infrastructure). Despite extensively used in emergency context, it offers a narrow bandwidth for ProeTEX perspectives and therefore it was discarded. *Wi-MAX* (Worldwide Interoperability for Microwave Access) is used for broadband wireless access in metropolitan areas, providing high data rate in line of sight conditions; there is a plan of supporting Wi-MAX in high and low mobility conditions too; nevertheless it is not a so mature technology and it is difficult to have portable application; in terms of power consumption, devices are far from optimality at the moment and frequency allocation (among countries) is still in a work-in-progress phase. *Wi-Fi* is an IEEE802.11 standard technology, born to extend LAN to wireless environment: it was conceived to be easily deployable, implementable in laptops and PDAs, low power consuming and performing on limited coverage area (typically in public spaces like hotels, airports, campus). At now, flexibility and miniaturization of the hardware make it a good compromise, offering link capacity (for services beyond the data monitoring), well-engineered devices (preserving battery supply), and high security level (no links of unauthorized users). Recent works demonstrated the suitability of 802.11.g/n protocols to set up ad-hoc wireless networks during fire-fighting interventions for real-time data and video transmission [20, 25]. Therefore this final protocol was adopted as a solution for the ProeTEX wearable system.

### 3.3. Architecture

Based on Wi-Fi protocol, the LRS is conceptually made up of 3 types of stations: *operators* (OP), *re-transmitting* (RT) and *remote* (RM) *nodes*. Each OP node is embedded within the OG and connected to PEB through a serial RS232 bus: it receives data strings containing the measurements gathered from all the ProeTEX sensors and remotely transmits these data. The RT nodes - provided with directional antennas and placed on tripods on the operative area - receive data from several OP nodes and forward them to the RM node (Figure 2). This latter one acts as an access point and is set at the margin of the operative area where coordinators supervise interventions of first line responders. In an operative scenario, it may be mounted on vehicles, i.e. the "local command post"; the RM node is provided with a notebook or PC running customized *remote monitoring software*, which displays data. An overview of the nodes' set-up is reported in the Figure 3 (see also [26]).

All 3nodes are based on the same hardware platform, an ALIX3c2 single board computer [27], equipped with 500 MHz AMD Geode LX CPU, 256 Megabytes of RAM, 2 mini-PCI slots, one RS232 serial port and a 1 Gb compact flash. Communications are performed by an integrated Ethernet port and a Wi-Fi card [28]: card operates in the 2.4 GHz band, requesting a maximum 360 mA current in the IEEE 802.11 b or g mode. To minimize power consumption and weight, hardware is powered by a dual cell Li-Po battery (7.4 Volts, 2100 mAh, 104 gr - MP157130 by Multiplex) and ensuring an autonomy of more than 4 hours. All components are packed into an aluminium enclosure (113x163x63 mm) embedded in the OG, within a customized pocket.

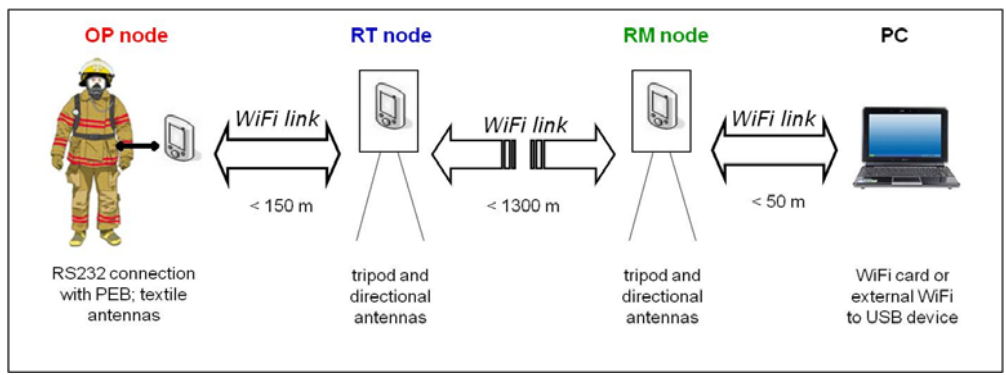

Figure 3. Architecture of the Long Range communication System: from left to the right, the operator (OP), re-transmitting (RT) and remote monitoring (RM) nodes respectively.

#### 3.3.1. OP node – mobile BAN

From a functional viewpoint the most remarkable node is the OP one: it boots from the compact flash a Linux distribution (Voyage Linux 0.5); the operating system is stored within the RAM, where two applications run: the 1$^{st}$ application polls PEB (according to a sampling time defined by the remote software user) and stores the obtained strings in a database, after tagging the data with sequential numbers; the 2$^{nd}$ application acts as a concurrent server waiting on a TCP port for the remote software's requests, i.e. sampling frequency modification, query update or data retrieval; this one is not a stream server, since the client does not subscribe an information flow, rather sends the sequential number of the last sample received; the server extracts from a database the next strings (up to 50) in order to deliver them to the client. This strategy makes the system asynchronous twice: data are locally saved in case of temporary connection problems and the receiving side can tailor link capacity on the basis of the processing needs. The most exciting feature is a real time monitoring of rescuers and not long sessions of data recording with subsequent bulk transfer.

Given the limited hardware resources, a SQLite system is in charge of database management; unfortunately it does not handle multiple accesses since it is a library. According to the aforementioned description, at least two independent processes have to deal with the database; therefore an external mechanism should be added: for efficiency reason the two applications are written in C and the native solution is represented by semaphores. The same applies to rule the access to shared memory area where common variables - used by the serial reader but changed on the client side - are stored (for example the sampling time or the query string). An overview of this OP data flow is detailed in Figure 4.

Finally, the wireless communications of the OP node rely on 2 textile antennas placed within the OG - in the front and back side respectively - and made of high-conductivity electro-textile material lying on flexible foam and shaped in a rectangular ring micro-strip patch [29].

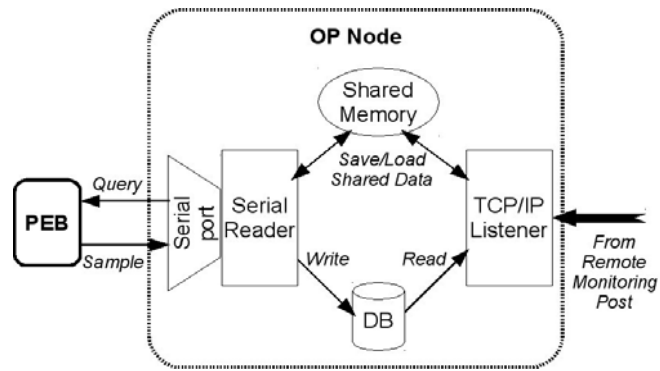

Figure 4. The OP node internal software view.

### 3.3.2. RT node – router

A bidirectional flow, covering the area from the OP to the RM nodes, requires two antennas with similar gain: textile technologies can not provide these performances; therefore a re-transmitting (RT) module was introduced. Once again, it is based on an ALIX board, extending the access point coverage by a Wireless Distribution System (WDS) mechanism. It is linked to the OP node thanks to an omni-directional antenna, while the one employed for the long range link toward the RM station is a 15 dB gain device with a 3 dB pattern equal to 30° by 30°. Since the RT module should not be worn and it should operate for a longer time, it can be powered by external and heavier battery. Moreover, for indoor and laboratory tests - where the involved distances are not significant - it is always possible to use OP and RM nodes only.

### 3.3.3. RM node – remote monitoring post and monitoring software

The remote monitoring (RM) post represents the terminal collecting data recorded by instruments worn by several first line responders. According to emergency practice, it should be placed in a safe place and made of easily transportable components. The RM node is composed by an embedded single board computer, identical to the ones of the former nodes, and by a notebook or PC. Two antennas are installed on the node: an omni-directional one assuring the Wi-Fi local connection with the PC, whilst communications with the RT node are provided by a 15 dB gain directional (90° by 15° in the horizontal and vertical plane, respectively) antenna. The PC runs customized software, which real-time displays each operator position plus data of all wearable sensors. According to end-user requests, the software should provide intuitive and simple information, since displaying signals of several operators and sensors may be useless and confusing; thus, the software was conceived to report relevant information in graphic format only. It was realized with object oriented VB.Net programming language in three components: a connection module, a data processing module and a graphic user interface.

The *connection module* manages up to 3 OP nodes, generates queries and receives answers, detects possible losses of connection, implements automatic reconnection routines and notifies the connection status.

The *data processing module* interprets each sensor data by real-time extracting information (and warnings) with routines based on comparison with tuned thresholds: it outputs warning codes, which are displayed on the graphic interface by coloured flags. Results of this real-time single sensor analysis are clustered to generate "global warning" flags about the whole health state of the user (defined by heart rate, breathing rate, body temperature, heat flux through the Outer Garment, activity and possible falls to the ground) and the surrounding environment (external temperature, toxic gases concentration). A third global "equipment state" flag is also managed, by taking into account the connection availability and batteries levels of the electronics.

Finally, the *graphical interface* shows the real time position of each operator on a map of the intervention area (interfaced with Google Earth software, provided by Google, Inc), thanks to the GPS module within the ProeTEX equipment. Each user is represented by an icon, whose colour changes according to the global "user health state" and "environment state": a grey colour is associated to losses of connection or unavailable GPS signal, green colour refers to normal user status, red colour means that one or more detected variables are outside the normality range. Figure 5 reports a screenshot of the interface.

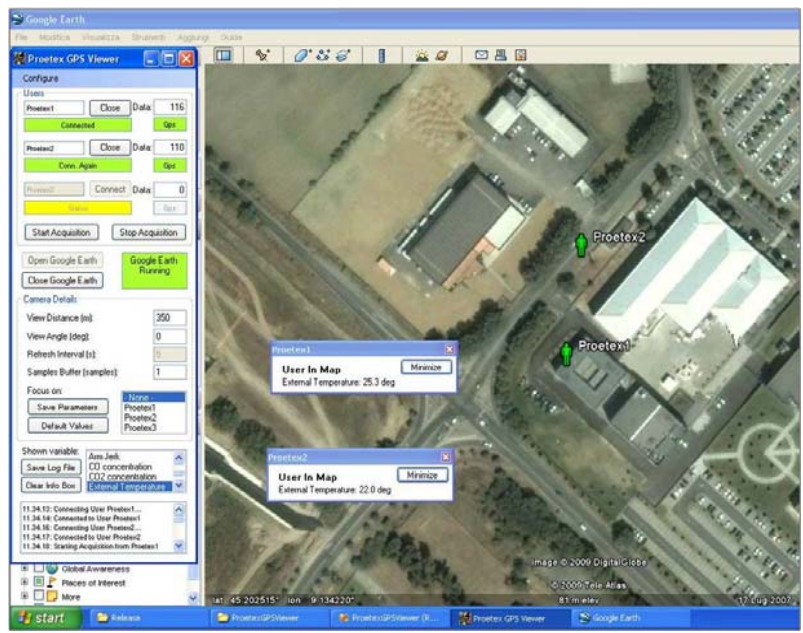

Figure 5. Screenshot of the ProeTEX operation field monitoring software coordinator.

## 4. SYSTEM VALIDATION

### 4.1. Experimental set-up

System's performance were proven in both short and long range conditions, according to the following guidelines: *short range* tests involved two nodes only, namely the OP and RM ones. The tester - wearing the ProeTEX equipment (i.e. the OP node) - was asked to walk for a distance of about 400 m from the RM node, then to turn back and walk toward the base-station (RM), in an obstacle-free and outdoor environment. During all the time, the monitoring software recorded the GPS position and sensors' data. Given this protocol, both the textile antennas - in the front and back sides of the OG - were tested. The experiment was repeated 4 times by using the omni-directional antenna on the RM node (case 1) and 4 more times with the directional antenna (case 2), in order to check the connection between the OP and RT nodes in field conditions and the gain of the textile directional antennas respectively. The *long range* tests envisaged all the 3 nodes (OP, RT & RM): these tests were focused on verifying the performances related to the link between RM and RT stations, while the OP node was kept close to this latter one. These conditions aimed at simulating a typical usage scenario, during which RM and RT nodes are set-up in specific places - before the operators start intervening - and they removed once the intervention has been concluded. Therefore, the monitoring was not continuous, rather the RT module was placed in 11 specific sites at increasing distances from the RM unit, and then a 10 min acquisition session was started in each site, in order to test the connection reliability during all the time.

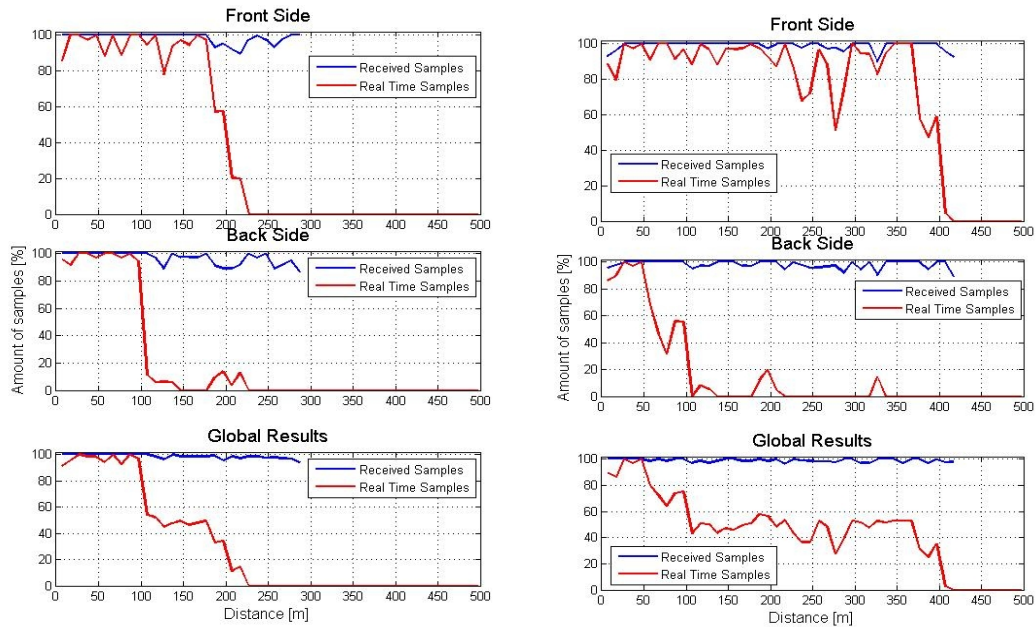

Figure 6. Performance of the communication system in short range configuration, with the *omni-directional antenna* (left side) and the *directional antenna* (right side) installed on the *RM node*: amount of correctly received samples (blue lines) and real-time received samples (red lines) when exposing the front side (upper panels) and back side (central panels) of the user garment to the RM node. Lower panels report the average results in the both the conditions.

### 4.2 Data analysis

Both tests aimed at evaluating the number of samples received without errors and - among them - the statistic of those received in real time, as a function of the distance between the transmitting and receiving nodes. Distance was evaluated exploiting the output of the GPS module and associated by PEB to each sample transmitted to the OP node. Precisely, the initial position of the motionless stations was measured by asking the subject to stay motionless for 2 min near the tripod hosting the node and by averaging the so recorded Latitude and Longitude values. Reliability of GPS measures was strengthened by the low Horizontal Dilution of Precision (HDOP - always below 1.5) and the high number of visible satellites (always more than 7).

Percentage of samples received without errors was measured as the ratio between the number of received samples - passing the parity check - and the number of expected samples - obtained by measuring the beginning and end time of each acquisition with PEB generating outputs at 1 Hz rate. Percentage of real-time received samples was measured by taking into account both PEB and monitoring software timestamps. In fact, PEB provided a timestamp, pointing out the time of its answer to the OP node. Similarly, the monitoring software associated a timestamp (in ms since midnight) to each received valid sample. Once removed an initial offset between the two values, a delay higher than 1 s between these timestamps proved that the off-body transmission was not real-time implemented. Finally, percentage of real-time received samples was computed as the ratio between the number of the remaining samples and the number of expected ones.

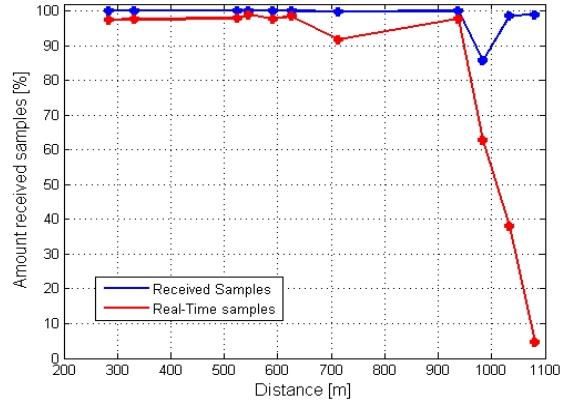

Figure 7. Performance of the communication system in the long range configuration, with *flat directional antennas* installed on the *RT* and on the *RM nodes*: amount of correctly received samples (blue line) and amount of samples received in real-time (red line) as a function of the distance between the nodes.

### 4.3 Results

Concerning the *short range* experiments, a total amount of 4979 samples was collected during the 2 sessions, over an expected value of 5039 (98.81 % of correctly received samples). No notice significant differences were noticed by using directional and omni-directional antennas on the RM node: in the former experiment the monitoring software correctly received 2644 samples over 2672 expected (98.95 %), while in the latter it received 2335 samples over 2367 (98.65 %). Despites almost all the samples were properly received, the tests showed differences within the percentages of the real-time transmissions, depending on distance between nodes and the experimental setup: left graphs of Figure 6 reports 3 plots displaying percentages of the correctly transmitted samples (blue lines) and the real-time transmitted ones (red lines) as a function of the distance between the OP and RM nodes, when an omni-directional antenna was used on the RM node. The right figure side shows the results of a similar analysis, carried out on data recorded when using a flat directional antenna on the same node. Results obtained when the user was moving away from the RM node (i.e. exposing the back side to the base-station) and approaching the node (i.e. exposing the front side) were noticeably different: upper panels of Figure 6 refer to the first case, whereas central panels refer to the second condition; finally, lower panels report the average results. Graphs show a large difference in terms of real-time transmission with the different configurations: this behaviour justifies the development of a data storage procedure in the OP node; in fact a deferred transmission - as soon as a stable wireless link is available – preserves the network from temporary losses of data due to displacements of the operator with respect to the base-station. Furthermore, these results presented refer to tests carried out in line-of-sight conditions: in operative scenarios, the presence of thick vegetation and obstacles may reduce the connection distance, thus requiring this deferred data transfer.

Concerning the *long range* experimental setup, the transmission performance was verified during 11 sessions, carried out by placing the RT and RM nodes at distances ranging from 280 to 1081 m. A total amount of 6045 samples was acquired, over an expected number of 6149 (98.31 %). No significant differences in the percentage of correctly received samples were noticed during the sessions. When nodes were placed at a distance lower than 900 m in open space, the amount of real-time transmitted samples was high and stable, whereas reliability of decreased at higher distances. Figure 7 summarizes the results, in terms of amount of correctly received samples (blue line) and real-time received samples (red line), as a function of the distance between the nodes: asterisks report the positions where measures were taken.

## 4. DISCUSSION AND CONCLUSION

This work describes a transmission system realized to meet guidelines arose during the ProeTEX project. The final goal of this project is a complete set of garments overseeing the health status of an emergency operator and possible dangerous environmental conditions. A resilient, light and wearable communication system - ensuring continuous rescuers surveillance together with suitable ranges preserving the team coordinator safety – has been conceived within the project too. The previous paragraphs gave details about the architecture and testing of this 3 nodes-based architecture: briefly, data are locally processed, transmitted and converted in easy understandable information within a graphical interface. The interface is usable by not medical personnel to coordinate emergency operations and to easily recognize critical situations.

In our opinion, the system developed in the ProeTEX project compares favourably with the state of the art [16, 17 and 25]. We adopted the Wi-Fi protocol as many other solutions in order to have an easy deployable network. We kept the architecture simple and without turning to ad hoc routing, the use of directional antennae was enough to comply with our constraints (as demonstrated in many field trials). In case the scenario gets complicated, involving more indoor areas, it is always possible to increase the number of retransmitting modules (the same as adding ad hoc routers suggested in [25]). Moreover, we handle link failures or outages simply by adding buffering on the wireless device. Finally, we integrated two aspects in the same interface: an environmental monitoring, to look for dangerous gases or other hazards, combined with preservation of operator's health conditions.

During experimental activities - aimed at testing the efficacy of the communication system - harsh conditions consisted in moving the relay node on different positions for each taken measure, every time building from scratch the experimental set-up. Even in a rural scenario, when distance was greater than 500 m, precise antennas pointing looked not so trivial, since human sight did not allow a clear view. Nevertheless, referring to practical cases, a rescue team should be able to position RT modules and get an immediate connection roughly driving the device towards the RM nodes, without wasting time. It is important to notice that this is not the case of a fixed sensor network, like in [30]: in this situation it will be reasonable to spend time to tune the deployment since - once the sensors are placed - they are not going to be moved for a long time. In the present case, in the trade-off between performances and ease of usage, the second aspect should be preferred; to achieve this goal we tested different directional antennas: finally, the radiation pattern combination performing best results is the one already proposed in par. 3.3 *Architecture*. Moreover this combination permits a manual orientation - without requiring maps or additional tools – and it allows a quick network set-up within the areas of interest.

Health monitoring platforms, as the one proposed in [31], rely on existing telecommunication infrastructures like cellular networks or public available Internet accesses. In this case, the idea is to exploit the metropolitan area network to connect closest hospitals to the victims. On the contrary, the ProeTEX communication system is self-consistent, and needs the aforementioned 3 nodes only. Our aim is to monitor a well-defined and limited area hit by the disaster (for example an earthquake, a fire, flooding, etc.), while the coordination centre stands in the proximity. Extending the range of the system is quite simple: it would be enough to place the RM modules on vehicles supporting a satellite connection; a similar solution has been already investigated in other projects [32].

The developed system has demonstrated its suitability in avoiding losses of data when the communication link between the operators and the local command post fails for short periods of time, due to user's displacement or presence of obstacles between the OP and RT nodes. Future actions will be devoted to enhance the real-time performance, reducing possible gaps in the data flow. From one point of view, we are monitoring some parameters that can potentially compromise the life of rescuers, therefore best accuracy and time continuity should be reached. However, a dangerous situation is not triggered by an abrupt change in readings: for example, CO concentration is fatal after high exposition for 15 min. If the communication fails for few seconds, all measurements can be saved locally and retrieved in a short interval when the connection is back again. Even with a small delay, supervisors have sufficient information to take decisions. The possibility for each OP node to act as a relay for the other operators or the introduction of a mesh network instead of single RT nodes could help in reducing these drawbacks.

Finally, one more aspect to be improved within the manufacturing of this system relies on miniaturization and integration. There is no need to have an electronic box (the PEB) to gather data from sensors and another device for transmission (the OP node). The solution proposed here is a prototype only: at present, the communication box was kept as a stand-alone unit in order to make all necessary changes in the software and network technology. Once the result is satisfactory, the power consumption and ergonomics could benefit from integration.

## ACKNOWLEDGEMENTS

This work is supported by the European Community Framework Programme VI, IST Programme, Contract n.26987.

**Authors**

**Matteo Lanati** received the MSc in Electronic Engineering from the University of Pavia in 2004 and the PhD from the same institution in 2007. Since then he started to collaborate with Eucentre (European centre for Training and Research in Earthquake Engineering) working on two European projects, ProeTEX (Protection e-Textiles: MicroNanoStructured fibre systems for Emergency-Disaster Wear), in order to develop a long range Wi-Fi based communication system, and DORII (Deployment of Remote Instrumentation Infrastructure), to port the selected seismic applications to the Grid infrastructure. 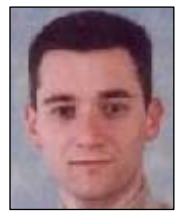

**Davide CURONE** got his laurea degree cum laude in Biomedical Engineering in 2005 and a PhD in Bioengineering and Bioinformatics in 2010, both at the University of Pavia (Italy). Since 2006 he has been working with the Technological Innovation research group of Eucentre, on topics concerning communication, signal processing and software development, within the EU-ICT IP ProeTEX. 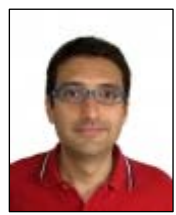

**Emanuele L. SECCO** is a postdoctoral research scientist at Eucentre, Pavia, Italy. He received the degree in Mechanical Engineering from the University of Padova in 1998. In 2001 he got a PhD in Bioengineering and Medical Computer Science at the University of Pavia and a Post-doc at the Rehabilitation Institute of Chicago in 2004. Since 2007 he has been working within the Innovation Technology team at Eucentre. Dr. Secco has been involved in bio-mimetic models, motor learning and rehabilitation technologies. Recently (2007-2010) he has been working on definition of specifications and requirements, testing and validation of smart garment prototypes for emergency context within the UE-ICT IP ProeTEX. 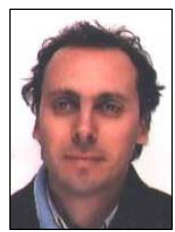

**Giovanni MAGENES** got his laurea degree cum laude in Electronic Engineering in 1981. In 1988 he received the PhD in Biomedical Engineering. Since 2005 he is full professor of Biomedical Signal Processing at the University of Pavia, Faculty of Engineering. He is the Director of the Department of Computer and System Science of the University of Pavia. At Eucentre he is leader of the Technological Innovation Section. Giovanni Magenes is author of more than 140 international publications. Recently Giovanni Magenes has been leader of the research unit at Eucentre in the UE project ProeTEX (FP6) and member of the research unit at Eucentre in the project STEP (FP6). 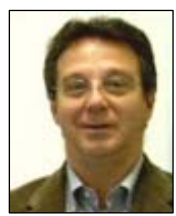

**Paolo Gamba** is currently Associate Professor of Telecommunications at the University of Pavia, Italy. He is also in charge of the Telecommunications and Remote Sensing Section of Eucentre. He is a Senior Member of IEEE, and since January 2009 he serves as Editor-in-Chief of the IEEE Geo-science and Remote Sensing Letters. He serves as Technical Co-Chair of the 2010 IEEE Geo-science and Remote Sensing Symposium, Honolulu, Hawaii, July 2010. His main research areas are the design of algorithms for airborne and spaceborne remotely sensed data interpretation for environmental monitoring and urban remote sensing, wireless systems and network design, and sensor networks. 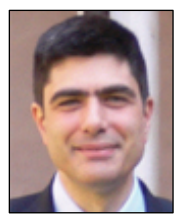